\documentclass{article}
\usepackage{spconf,  amsmath,  graphicx}
\usepackage[colorlinks,  citecolor=blue,  urlcolor=blue,  bookmarks=false,  hypertexnames=true]{hyperref}
\usepackage{bbm}
\usepackage{dsfont}

\title{End-to-end speaker diarization with transformer}

 \name{Yongquan Lai,  Xin Tang*\thanks{*Corresponding author},  Yuanyuan Fu,  Rui Fang}
\address{Visual Computing Group,  Ping An Property \& Casualty Insurance Company,  Shenzhen,  China}
\begin{document}
%
\maketitle
\begin{abstract}
Speaker diarization is connected to semantic segmentation in computer vision.   Inspired from MaskFormer  \cite{cheng2021per} which treats semantic segmentation as a set-prediction problem,   we propose an end-to-end approach to predict a set of targets consisting of binary masks,  vocal activities and speaker vectors.
Our model,   which we coin \textit{DiFormer},   is mainly based on a speaker encoder and a feature pyramid network (FPN) module  to extract multi-scale speaker features which are then fed into a transformer encoder-decoder to predict a set of diarization targets from learned query embedding.   To account for temporal characteristics of speech signal,   bidirectional LSTMs are inserted into the  mask prediction module to improve temporal consistency.  Our model handles unknown number of speakers,   speech overlaps,   as well as vocal activity detection in a unified way.  Experiments on multimedia and meeting datasets demonstrate the effectiveness of our approach.

\end{abstract}
\section{Introduction}
\label{sec:intro}
\begin{figure*}[!h]
	\centering
	\includegraphics[width=0.8\textwidth]{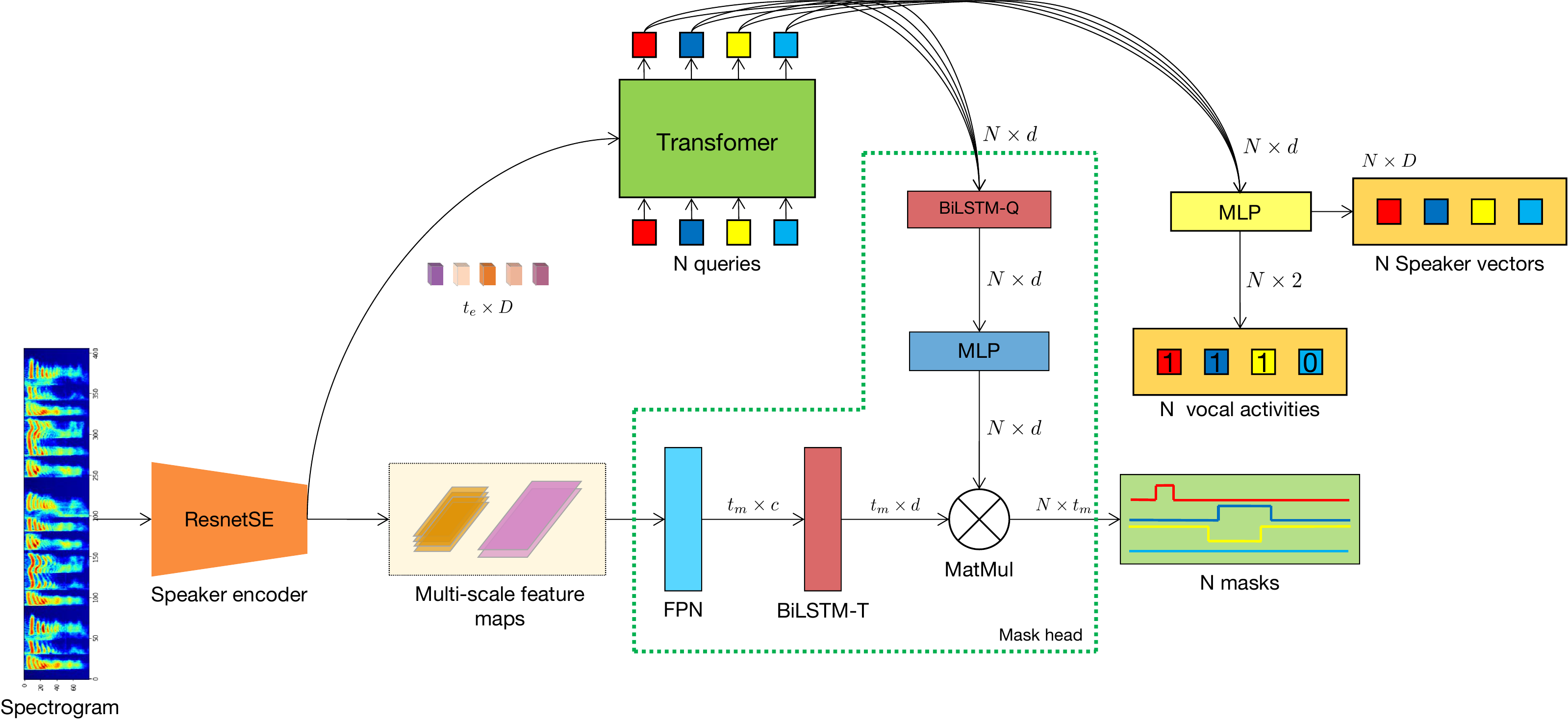}
	\caption{The architecture of proposed model mainly consists of speaker encoder,  transformer,   feature-pyramid network (FPN) and mask attention module.
	Best viewed in color.}
	\label{fig:arch}
\end{figure*}
Speaker diarization (SD) is the task of answering "who spoke when'' question,   given a single-channel audio that contains speech from multiple speakers.
It takes the audio signal as input and produces multiple 1d masks  along with corresponding speaker identity information.
Since 1d masking is a special case of  2d masking,   connections can be drawn  from semantic  segmentation in computer vision to speaker diarization in audio,   especially when audio signal is represented by spect-rogram.

In semantic segmentation,   usually we assume there are a finite set of semantic labels,   which can be predicted using  a fixed class head.  However,  for the SD problem,   the speaker identities that appeared in the recordings are unknown.  Therefore it is impossible to train a classifier to the output class label for each identity,   given the infinitely many distinct speakers in the world.   Another feature of SD is that the signal to process  is a temporal signal. Hence temporal consistency for speaker mask and speaker identity should be considered.

Our work is highly motivated from MaskFormer \cite{cheng2021per},
which is an elegant approach to solving semantic segmentation problem by predicting a set of binary masks and semantic labels. By drawing connections from semantic segmentation to SD,   we propose to solve the SD problem by modeling it as a  set  prediction problem.   Our model predicts an unordered set of targets,   each of which contains a 1d segmentation mask,   a speaker vector representing the speaker identity,  and a class probability for vocal activity  detection (VAD).   As shown in Fig.  \ref{fig:arch},   the audio signal is represented by spectrogram and temporal speaker features  are extracted by speaker encoder and fuse with a FPN. A transformer-based encoder-decoder module predicts a set of targets based on learnable queries. Two bidirectional LSTMs (BiLSTMs) are inserted to encourage temporal consistency for mask prediction,  making the prediction smoother.

Our model is innovated in multiple ways: 1) it is the first effort to address SD as a set-prediction problem,   2) speaker embedding prediction rather than class id is used to tackle the problem of unknown speaker number,  and 3) the introduction of BiLSTMs to assure smooth  binary mask prediction.  Our model can naturally handle speech overlapping since the mask regions are not exclusive.  It is also invariant to speaker permutation since the masks are treated as a (unordered) set.  It works well for unknown number of speakers and does not require estimating the number of speakers.


\textbf{Relative works} SD can be tackled by using a speaker encoder and a clustering algorithm to group together temporal features that belong to the same speaker, e.g.,  \cite{karra21_interspeech,landini2022bayesian}.  These methods usually involve multiple stages and require heavy parameter tuning.
The VBx mothod  \cite{landini2022bayesian} uses a Bayesian hidden Markov model for speaker clustering from a sequence of x-vectors and has achieved low diarization error rate (DER).  However,   the embedded and spectral clustering  approaches need to know the number of speakers beforehand,   either by estimation or manual setup.   The VBx method performs better if initialized by the classic agglomerative hierarchical clustering algorithm (AHC).  End-to-end diarization is an active research direction,  e.g.,  \cite{liu21j_interspeech,  maiti2021end}.  An overlap-aware segmentation and VAD approach is proposed in \cite{bredin21_interspeech} as a post-processing step for VBx. To address the identity swap problem,   permutation invariant training has been applied to speech separation  \cite{yu2017permutation, yang2020interrupted}.  LSTMs are popular in   end-to-end SD,  e.g.  \cite{fujita2019end, horiguchi2020end}.  Due to page limitation,   we refer readers to  \cite{park2021review} for a comprehensive  review of deep learning-based methods,   datasets,   metrics,   etc.


\section{DiFormer}

The proposed DiFormer model mainly consists of speaker encoder,   transformer encoder-decoder,   mask prediction head,   VAD head and speaker head.  The speaker encoder extracts temporal speaker information,   which will be processed by transformer and mask prediction module to generate $N$ mask alignment vectors.
The alignment vectors later play the role of generating $N$ masks by attending to $t$ feature vectors with sigmoid activation,   resulting in  $N$ $t_m$-dimensional binary masks.  The query-based transformer encoder-decoder takes the length-$t$ speaker features as  input,    and produces a length-$N$ feature sequence that encodes global speaker and speech information. The processed feature sequence is useful for predicting binary speaker mask,   as well as the speaker identity vectors and vocal activity indicators.  We now introduce each module in details.

\textbf{Speaker encoder} is a pretrained Resnet  \cite{he2016deep} with squeeze-and-excitation module  \cite{hu2018squeeze} (ResnetSE).
The single-channel audio waveform is converted to time-frequency domain before feeding into speaker encoder.  We follow the convention of using log-mel spectrogram  in speaker verification and identification tasks.  Details are described in section \ref{sec:imp}.

Within the whole training process,  the encoder is put in eval mode and no gradient is computed for its parameters. The speaker encoder originally produces output vector of $D$-dimensional $L_2$-normalized vector for speaker embedding. We modify the encoder to generate frame-level speaker embedding: a $t_e \times D$ tensor to feed into transformer module to further processing,  where $t_e$ is sequence length of the embedding.

The speaker encoder has four layers and each of the last three layers halves mel-frequency and spatial dimension. The feature maps of last two blocks denoted as $\mathbf{x}_l$ and $\mathbf{x}_h$ are fed into the subsequent FPN for multi-scale fusion.

\textbf{Mask prediction head},  as shown inside the green dashed lines of Fig.  \ref{fig:arch},  consists of two BiLSTM,  namely a BiLSTM-Q for smoothing the length-$N$ Queried outputs,  and a BiLSTM-T for smoothing the multiscale Temporal feature sequence generated by FPN. The output of  BiLSTM-Q will be further processed by MLP,   as in  \cite{cheng2021per},   to generate alignment vectors $N \times d$ for attending to the $t_m \times d$ temporal features by matrix multiplication,   resulting in the final $N \times t_m$ mask prediction. $t_m$ is proportional to origin audio duration. Sigmoid activation is appended to the MLP and the final mask prediction.
To alleviate the over-fitting problem,   a $10\%$ dropout is added before both LSTMs (not shown in Fig.  \ref{fig:arch}).

\begin{figure}[htb]
      \centering
	\includegraphics[width=0.9\linewidth]{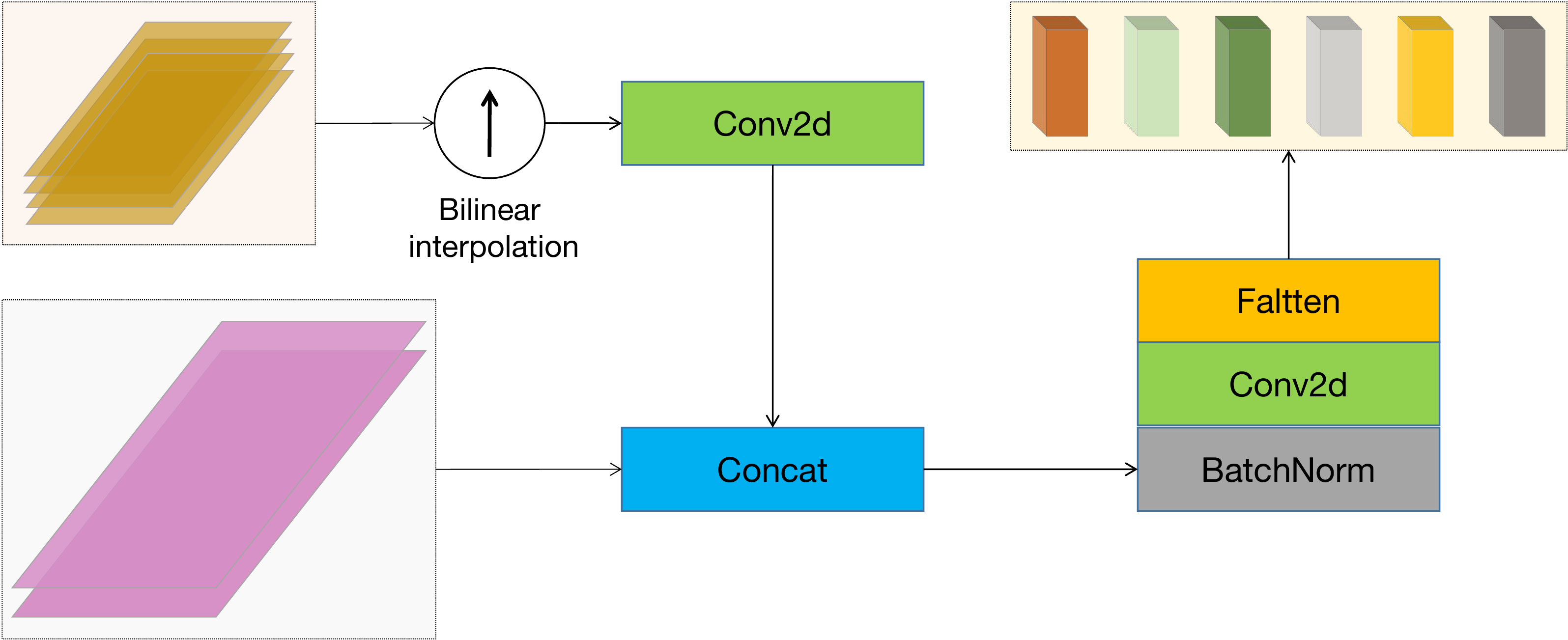}
	\caption{The FPN module up-samples low-res features  and combines them with high-res features. This will increase diarization resolution while still retain high-level information.}
	\label{fig:fpn}
\end{figure}

The feature-pyramid network (FPN) module,  as depicted in details in Fig.  \ref{fig:fpn},   accepts low/high resolution feature maps $\mathbf{x}_l$/$\mathbf{x}_h$ and generates a tensor of  shape $t_m \times c$.  Mathematically,   the FPN implements $\mathbf{x}=\text{conv}(\text{bn}(\text{cat}[\text{conv}(\mathbf{x}_l\uparrow),  \mathbf{x}_h]))$.

\textbf{VAD head} is implemented as a MLP with softmax activation.  It predicts $N$ probability distributions independently,   each representing how likely does the query contain meaningful speech signal.  The groundtruth VAD label $\mathbf{v}_j^{gt}$is crucial for optimization.  For $\mathbf{v}_j^{gt}=1$,   the $j$th query of speaker embedding and mask predictions is not  compared against the corresponding groundtruth.  In other words,   the masks for non-vocal signal segments are not forced to be zero,   so is the speaker embedding.

\textbf{Transformer module} consists of transformer encoder and decoder which are nearly identical to  \cite{Attention} except that pre-normalization is used.  Both encoder and decoder have three layers,   and fixed cosine positional embedding is added.  The decoder uses $N$ learnable query embedding.

\textbf{Multitask optimization} The outputs of the DiFormer model consist of three parts,   namely $N$ binary mask vectors,    $N$ indicators for vocal activities and $N$  predicted speaker embedding vectors $\in R^{D}$.  The speech indicators play the role of vocal activity detection. They are implemented by $N$ two-class classifiers and optimized by cross-entropy loss with softmax activation.  The masks and speaker embedding are only meaningful if the corresponding speech indicator outputs $1$.  Overall,   the model predicts a set of masks and speaker vectors.  To measure discrepancy between prediction set and groundtruth set,   bipartite matching  is adopted to solve
the assignment problem,   as described in  \cite{cheng2021per}.  Our approach can be interpreted mathematically as follows.
Let $\mathbf{z}=\{\mathbf{v},  \mathbf{m}, \mathbf{e}\}$ denote the prediction of vocal activities,  masks and speaker embedding respectively.  $\mathbf{v} \in \{0,  1\}^N$,  $ \mathbf{m}\in {0,  1}^{N\times t_m}$,  $\mathbf{e} \in R^{N\times D}$,    $||\mathbf{e}_j||_2=1$ for $j=1,  ...,  N$.  $\mathbf{z}^{gt} = \{\mathbf{v}^{gt},  \mathbf{m}^{gt},  \mathbf{e}^{gt}\}$ denote the groundtruth set. Then the loss is defined as follows,
\begin{equation}
\begin{aligned}
\mathcal{L}(\mathbf{z}, \mathbf{z}^{gt}) &=\sum_j^N[P_{\sigma(j)}(\mathbf{v}_j^{gt})+\alpha\mathbbm{1}_{\mathbf{v}_j^{gt}=1}\mathcal{L}_{mask}(\mathbf{m}_{\sigma(j)},  \ \mathbf{m}_j^{ct})\\
&-\beta\mathbbm{1}_{\mathbf{v}_j^{gt}=1} 	\langle  \mathbf{e}_{\sigma(j)},  \ \mathbf{e}_j^{ct}	\rangle ]
\end{aligned}
\end{equation}
where $\alpha$ and $\beta$  are balancing weights. $\langle,  \rangle $ denotes inner product,   which actually computes the cosine similarity between groundtruth speaker embedding and the prediction.  Note that the ground truth speaker embedding vectors are pre-computed from the audio (with other speakers masked out),   by using the same pretrained speaker encoder.

\section{Experiments}
\label{sec:imp}
We evaluate our model using free opensource datasets,   namely VoxConverse  \cite{chung2020spot} and AMI corpus  \cite{mccowan2005ami}.  These two datasets are both large and hard enough for evaluating diarization models.

\begin{figure}[htb]
      \centering
	\includegraphics[width=\linewidth]{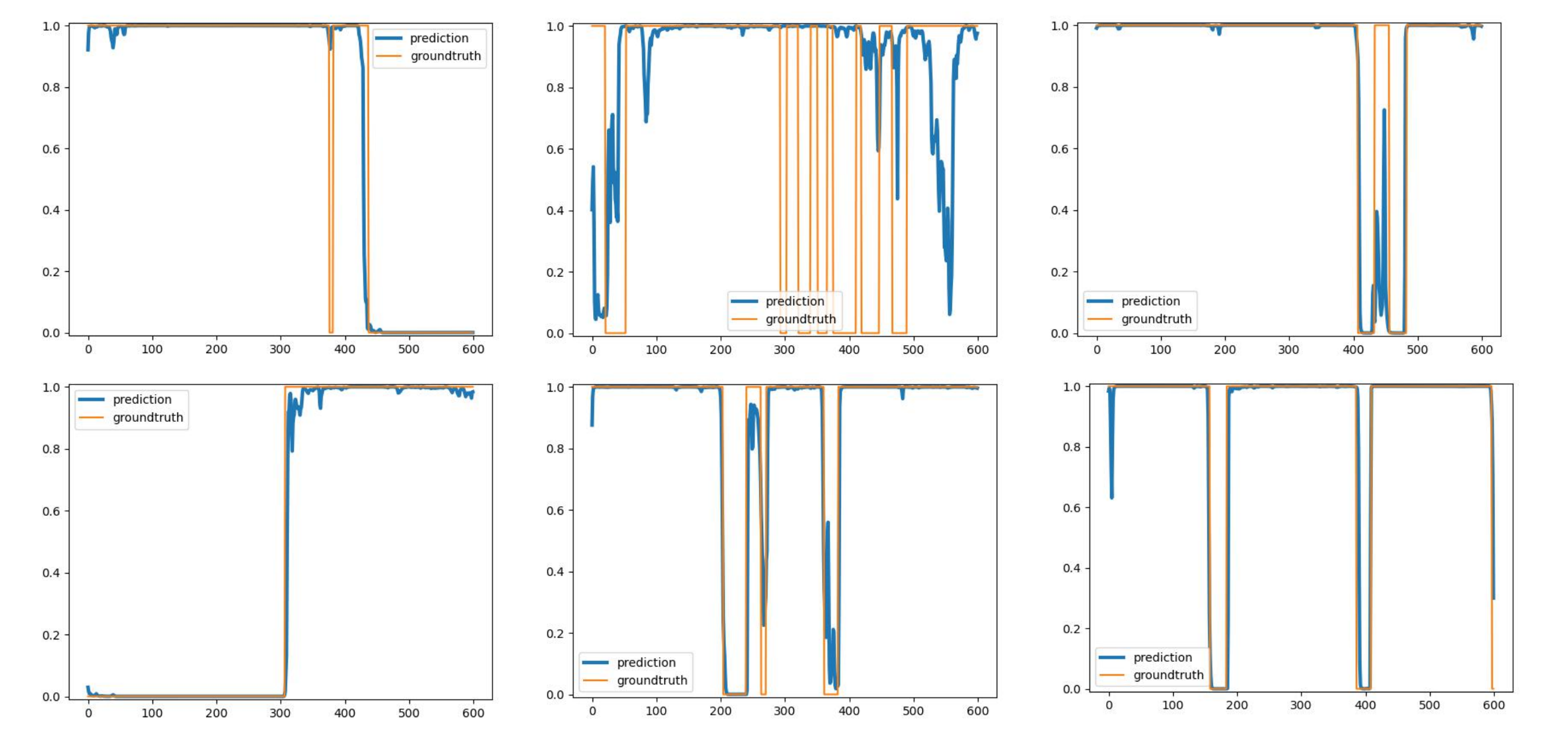}
	\caption{The first row shows some hard speaker turns  that our model fails to capture .  The second row shows successful cases where the boundary is accurately tracked.}
	\label{fig:case1}
\end{figure}

\textbf{Implementation details} The audio is converted to a log-mel spectrogram with parameters consistent with those used in training speaker encoder: sr=16000,   n\_fft=512,  win\_length=25ms,  hop\_length=10ms,
 n\_mels=80,   amin=1e-5,   top\_db=75.  Under this setting,  the downsample factor from original audio to spectr-ogram feature is $160$.  We use torchaudio to extract spectrogram features and the model is implemented in pytorch.  It is critical that the sound waves are normalized to [-1,  1] before  feature extraction.  We use the speaker encoder \footnote{https://github.com/ranchlai/speaker-verification} pretrained from VoxCeleb1\&2 datasets  \cite{nagrani2017VoxCeleb,  nagrani2020VoxCeleb}.

In all our experiments,  we assume that all audio recordings contain no more than 22 speakers and set $N=22$.  This is a safe guess for many cases including VoxConverse and AMI.  The model dimension $d$ in Fig.  \ref{fig:arch} is set to 512 and speaker dimension $D$ is $256$.  For input to FPN,  $x_h$ is $128\times 20 \times t_m$ while $x_l$ is $256\times 10 \times \frac{t_m}{2}$,  where $t_m$ equals to original audio length divided by 640,   or the frame length of spectrogram divided by 4.  After FPN,   the shape of  output feature map is $128\times 20 \times t_m$ and will be flattened and permuted to $t_m \times 2560$.
In training,   we set audio wave length of
2 (12 seconds). The loss balancing weights are set to $\alpha=1.0$ and $\beta=0.1$ to compensate scale difference.
 The number of parameters of the model is about 46 millions including the pretrained speaker encoder.  The speaker encoder has 6 millions parameters and they are fixed during all experiments.

As for many recordings,   especially in meeting scenarios,   the audio duration can be as long as tens of minutes.  One has to divide it into smaller segments and later combine the results back.   This is even true for end-to-end approaches,   as the modeling context and the GPU memory are both limited.  To deal with such recordings,   we use a simple sequential greedy clustering to achieve this: for each newly detected speech signal (or mask),   assign it to previous found identity if the speech is similar as measured by the pretrained speaker encoder,   otherwise a new speaker id is assigned.  The similarity threshold is simply set to  0.4 for VoxConverse and AMI.  Optimizing this threshold may gain performance.  Note that the clustering here is high level in the sense that it is used only to group speakers from different recording segments.

The speaker encoder is pretrained on VoxCeleb1\&2 datasets while the other modules in DiFormer are initialized randomly or pretrained from synthetic dataset.  We hypothesize that the later one will generalize better.  Hence we compose a synthetic training set from dev split of VoxCeleb1 and use the test split for evaluation.
Tab.    \ref{table:main} verifies our hypothesis:  pretraining on synthetic data reduces about 0.6 DER.

\textbf{VoxConverse} is a free multimedia dataset collected from Youtube semi-automatically. We use the latest official dev/test splits and compare some variants of our model with previous state of the arts (except  \cite{bredin21_interspeech} which uses different dataset version). When pretrained on VoxCeleb1 for 60K steps with four Tesla v100 GPUs and  finetuned on VoxConverse for another 100K steps (batch size  32 per GPU,   audio duration 12 seconds,    Adam is used with lr=1e-4 for pretrained and decrease to 1e-6 at every 50K steps at fine-tuning),   our model achieves lowest DER.

\label{sec:exp}

\begin{table}[h]
	\centering
	\setlength\tabcolsep{3pt}
	\begin{tabular}{|l|c|}
		\hline
		 Model &DER  \\
		\hline
		DIHARD 2019 baseline w/ SE  \cite{sell2018diarization} &20.2 \\
		\hline
		SyncNet ASD \cite{chung2019perfect,chung2020spot}&10.4\\
		\hline
		audio+visual  \cite{chung2020spot} &7.7\\
		\hline
		DiFormer (pretrained,   ignore overlaps) & 7.07 \\ 
		DiFormer (pretrained,   with overlaps) & 8.21 \\ 
		DiFormer (scratch,   ignore overlaps) & 7.68  \\ 
		DiFormer (scratch,   with overlaps) & 8.84\\ 
		\hline
	\end{tabular}
	\caption{Comparison with previous works on VoxConverse.  Our approach achieves competitive results using only audio stream.   Note that all previous methods ignore overlapping speech when computing DER,   making the problem easier.  When pretrained on simulated data form VoxCeleb1,   our model achieves lowest DER if overlapping is ignored.  Even when taking overlaps into account,   our model is still very competitive.}
	\label{table:main}
\end{table}

\begin{table}[h]
	\centering

	\setlength\tabcolsep{3pt}
	\begin{tabular}{|c|c|c|}
		\hline
		 Model&\multicolumn{2}{c|}{DER}\\
		 \hline
		 &Dev&Eval \\

		\hline
		Hierarchical clustering (AHC)  & 19.61 &21.43\\

		\hline
		VBx \cite{landini2022bayesian}&16.33 &18.99\\

		\hline
		Hybrid \cite{bredin21_interspeech} & - &19.9\\

		\hline
		2P-NB-LGP  \cite{karra21_interspeech} & 16.21 & 17.57\\

		\hline
		Overlap-aware  (Online) \cite{coria2021overlap}&-&27.5\\
		\hline
		DiFormer &  21.62 & 22.81 \\

		\hline
	\end{tabular}
	\caption{Comparison with previous works on AMI headset-mix. Protocol is the same as \cite{landini2022bayesian}.  Overlapping speech is NOT ignored,   collar is set to 0 (full setup).  All other approaches require sophisticated clustering while our model is end-to-end without sophisticated offline clustering and no heavy parameter tuning is needed.  Since we used a sequential speaker assignment algorithm for ultra long recordings,  our approach can be seen as an online algorithm with latency about the input duration,  i.e.,  12 seconds.}
	\label{exp:ami}
\end{table}

\begin{table}[h]
	\centering
	\setlength\tabcolsep{3pt}
	\begin{tabular}{|c|c|}
		\hline
		 Model &DER\\
		\hline
		Baseline &15.99\\
		\hline
		Baseline + BiLSTM-Q&12.10\\
		\hline
		Baseline + LSTM-Q\&T&9.61\\
		\hline
		Baseline + BiLSTM-Q\&T&8.45\\
		\hline
		Baseline + BiLSTM-Q\&T w/o FPN& 9.93\\

		\hline
		Baseline/1 layer transformer + BiLSTM-Q\&T&9.20\\

		\hline
	\end{tabular}
	\caption{Ablation study.   Each variant is trained from scratch (except speaker encoder) for 70K steps on VoxConverse.  Adding BiLSTM-Q decrease DER score  for 3.89,    and adding BiLSTM-T obtains additional 3.65 improvement,   verifying the importance of both module.  Moreover,  Bidirectional is better than directional.  FPN also contributes a lot.}
	\label{tab:abl}
\end{table}

\textbf{AMI dataset}  For AMI \cite{mccowan2005ami} we use headset-mix in this paper,  which contains 100 hours of speech.  We follow the setup here \footnote{https://github.com/BUTSpeechFIT/AMI-diarization-setup}.   The dataset is partitioned into train/dev/eval sets.  The training set contains about 70 hours of speech.  The dev and eval each contains 15.0/14.5 hours of speech respectively.  When overlapped speech is ignored,    the dev/eval sets reduce to  8.1/8.2 hours respectively.  The results are shown in Tab.  \ref{exp:ami}.  Our model,  although does not require sophisticated clustering and parameter tuning,   is able to perform nearly as well as previous works.   We also observe that pretraining from VoxCeleb1 is not as good as training from scratch.  This might due to that fact that
AMI is meeting dataset quit different form VoxCeleb.  In the future we will try using  libriTTS \cite{zen2019libritts}  for pretraining.

\textbf{Ablation study} is conducted to  the verify effectiveness of LSTMs in mask prediction module.  Starting from the baseline DiFormer model without LSTMs,   we subsequently add BiLSTMs or LSTMs and  train all variants fairly for the same steps from scratch using VoxConverse.   The result verifies our design,  as shown in Tab.  \ref{tab:abl}.

\section{Conclusions and future works}

We have proposed an end-to-end transformer-based model  to address  speaker diarization by treating it as a set-prediction problem.  Unlike clustering-based methods, our model does not involve multi-stage processing.   Experiments on AMI and VoxConverse datasets have verified the effectiveness of our model.
 To avoid  greedy sequential clustering in case of ultra-long sequence,   we can try to use the last hidden state of LSTM so that each segments are connected.   For meeting scenario such as AMI corpus,   pretraining on more clean and controlled datasets will help and we leave it for future works.


\bibliographystyle{IEEEbib}
\bibliography{strings,  diformer}
\end{document}